\documentclass[prd,preprint,superscriptaddress,showpacs,nofootinbib,%
tightenlines ]{revtex4}
\usepackage{graphics}
\usepackage{epsfig}
\newcommand{\ben}{\begin{displaymath}}
\newcommand{\een}{\end{displaymath}}
\newcommand{\be}{\begin{equation}}
\newcommand{\ee}{\end{equation}}
\newcommand{\bea}{\begin{eqnarray}}
\newcommand{\eea}{\end{eqnarray}}

\begin{document}
\preprint{MKPH-T-03-18}
\title{Infrared and extended on-mass-shell renormalization of two-loop
diagrams}
\author{Matthias R.~Schindler}
\affiliation{Institut f\"ur Kernphysik, Johannes
Gutenberg-Universit\"at, D-55099 Mainz, Germany}
\author{Jambul Gegelia}
\thanks{Alexander von Humboldt Research Fellow}
\affiliation{Institut f\"ur Kernphysik, Johannes
Gutenberg-Universit\"at, D-55099 Mainz, Germany}
\affiliation{High Energy Physics Institute,
Tbilisi State University,
University St.~9, 380086 Tbilisi, Georgia}
\author{Stefan Scherer}
\affiliation{Institut f\"ur Kernphysik, Johannes
Gutenberg-Universit\"at, D-55099 Mainz, Germany}
\begin{abstract}
   Using a toy model Lagrangian we demonstrate the application of both
infrared and extended on-mass-shell renormalization schemes
to multiloop diagrams by considering as an example a two-loop self-energy
diagram.
   We show that in both cases the renormalized diagrams satisfy a
straightforward power counting.
\end{abstract}
\date{October 17, 2003}
\maketitle
\section{Introduction\label{introduction}}
   The generalization of mesonic chiral perturbation theory
\cite{Weinberg:1979kz,Gasser:1984gg,Gasser:1984yg} to the one-nucleon sector
originally turned out to be problematic due to the fact that higher-order
loops contribute to lower-order calculations \cite{Gasser:1988rb}
(for a recent review see, e.g., Ref.\ \cite{Scherer:2002tk}).
   This problem was first overcome in the framework of the so-called
heavy-baryon  approach to chiral perturbation theory
\cite{Jenkins:1991jv,Bernard:1992qa}.
   More recently, it has been realized that the power counting can also be
restored in the original manifestly Lorentz-invariant formulation
\cite{Tang:1996ca,Ellis:1997kc,Becher:1999he,Gegelia:1999gf,Lutz:1999yr,%
Lutz:2001yb,Fuchs:2003qc,Fuchs:2003sh,Fuchs:2003kq}.
   In this context, the infrared (IR) regularization of Becher and Leutwyler
\cite{Becher:1999he} (based on ideas of Ref.~\cite{Tang:1996ca,Ellis:1997kc})
is the most widely used renormalization
scheme\footnote{Since the IR regularization is in
fact a renormalization scheme we will further on refer to it as
\emph{IR renormalization}.}
for one-loop diagrams
\cite{Ellis:1999jt,Kubis:2000zd,Zhu:2000zf,Kubis:2000aa,Goity:2001ny,%
Becher:2001hv,Borasoy:2001ik,Beisert:2001qb,Zhu:2002tn,Beisert:2002ad,%
Bernard:2002bs,Gasser:2002am,Torikoshi:2002bt,Bernard:2003xf,Higa:2003jk,%
Higa:2003sz}.
   In Ref.~\cite{Lehmann:2001xm} it has been suggested that the IR
renormalization can also be generalized to multiloop diagrams.
   In a recent paper \cite{Schindler:2003xv}, we have reformulated the
IR renormalization in a form analogous to the extended on-mass-shell
(EOMS) scheme of Ref.~\cite{Fuchs:2003qc}.
   In its new formulation the IR renormalization can also be directly applied
to diagrams involving resonances as well as multiloop diagrams.
   In Ref.~\cite{Gegelia:1999qt} the application of the EOMS scheme to
two-loop diagrams has been illustrated for the nucleon self-energy ``rainbow"
diagram in the chiral limit.
   The aim of this work is to demonstrate the application of the reformulated IR
renormalization \cite{Schindler:2003xv} to two-loop diagrams.
   For comparison the results of the EOMS renormalization are also given.

\section{Lagrangian and power counting\label{lpc}}
   In order to simplify the calculations and---without loss of generality---to
make the renormalization procedure more transparent, we suppress the
complicated spin and chiral structure of baryon chiral perturbation theory and
restrict ourselves to the toy model Lagrangian
\begin{equation}\label{lagrange}
\mathcal{L}=\frac{1}{2}(\partial_{\mu}\pi\partial^{\mu}\pi-M^2\pi^2)+
\frac{1}{2}(\partial_{\mu}\Psi\partial^{\mu}\Psi-m^2\Psi^2)-
\frac{g}{4}\pi^2\Psi^2+\mathcal{L}_1,
\end{equation}
where $\pi$ and $\Psi$ are scalar particles with masses $M$ and
$m$, respectively, $M\ll m$, and $g>0$.
   The Lagrangian $\mathcal{L}_1$ contains all interaction terms which are
consistent with Lorentz invariance and the invariance under the transformations
$\pi\to -\pi$ and $\Psi \to -\Psi$.\footnote{No particular physical content is
meant by imposing these discrete symmetries.} Both the IR and the EOMS
renormalization are constructed such that, after subtraction,
Feynman diagrams have certain "chiral" orders $D$, which are
determined by the following power counting: if $Q$ stands for
small quantities such as the $\pi$ mass $M$, small external four
momenta of $\pi$ or small external three-momenta of $\Psi$, then
the $\pi\Psi$ interaction explicitly shown in Eq.~(\ref{lagrange})
counts as $Q^0$, the $\Psi$ propagator as $Q^{-1}$, the $\pi$
propagator as $Q^{-2}$, and a loop integration in $n$ dimensions as
$Q^n$, respectively.

\section{Application to the two-loop self-energy\label{application}}
\begin{figure}
\epsfig{file=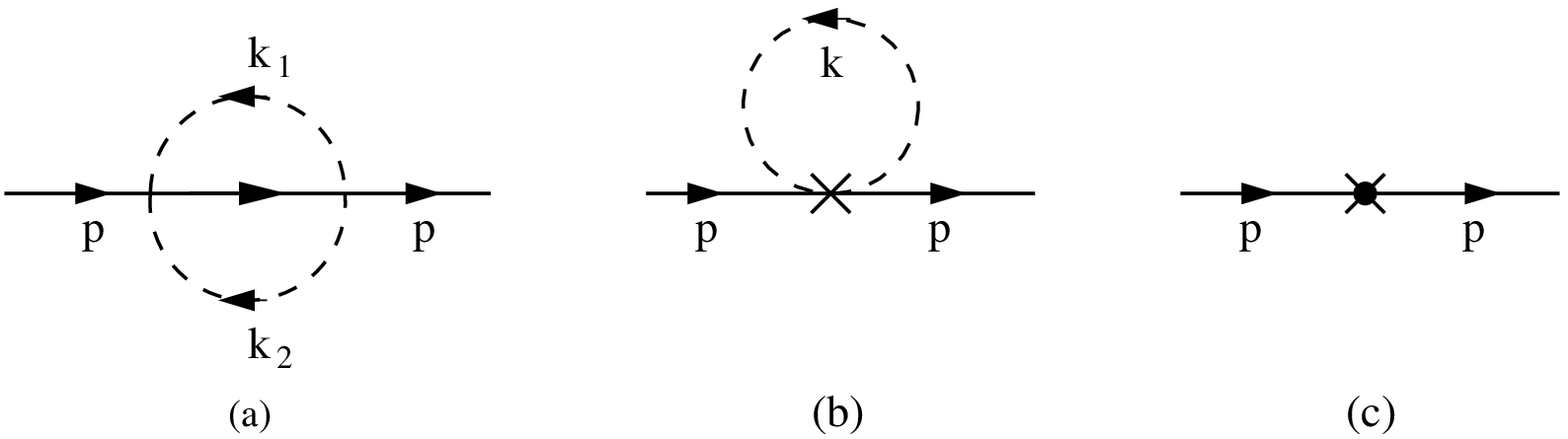,width=14cm} \caption[]{\label{fse:fig}
Diagrams contributing in two-loop order to the self-energy of
$\Psi$. A simple cross corresponds to one-loop order counterterms
and a cross with a dot corresponds to two-loop order
counterterms.}
\end{figure}

   As a specific example of a two-loop diagram, we consider the $\Psi$
self-energy diagram of Fig.~\ref{fse:fig}~(a),
\begin{equation}\label{selfenergy}
-i\Sigma_{\Psi}(p)=-\frac{i}{2}g^2J_{\pi\pi\Psi}(0,0,p),
\end{equation}
where $1/2$ is a symmetry factor and
\begin{equation}\label{orint}
J_{\pi\pi\Psi}(0,0,p)\equiv
\left(\frac{i}{(2\pi)^n}\right)^2\int\hspace{-2mm}\int
\frac{d^nk_1d^nk_2}
{(k_1^2-M^2+i0^+)(k_2^2-M^2+i0^+)[(p+k_1+k_2)^2-m^2+i0^+]}
\end{equation}
with $n$ denoting the number of space-time dimensions.
   Using the above power counting, we assign the order $Q^{2n-5}$ to the
diagram of Fig.~\ref{fse:fig} (a).
   A dimensional counting analysis \cite{Gegelia:zz} (see the Appendix
for an illustration) suggests that,
if first $M\rightarrow0$ and then $p^2-m^2\rightarrow0$,
the self-energy $\Sigma_{\Psi}$ can be written as
\begin{equation}\label{intsplit}
\Sigma_{\Psi}=F(p^2,m^2,M^2,n)
+M^{n-2}G(p^2,m^2,M^2,n)+M^{2n-4}H(p^2,m^2,M^2,n).
\end{equation}
   The functions $F(p^2,m^2,M^2,n)$, $G(p^2,m^2,M^2,n)$, and
$H(p^2,m^2,M^2,n)$ can be expanded in nonnegative integer powers of
$M^2$ and, in the following, will be analyzed in detail.

   First, we consider $F(p^2,m^2,M^2,n)$.
   The Taylor expansion of $F$ in $M^2$ is obtained by expanding the integrand
of Eq.~(\ref{orint}) in $M^2$ and interchanging summation and integration:
\begin{equation}\label{F}
F(p^2,m^2,M^2,n)=-\frac{g^2}{2}\sum_{i,j=0}^{\infty}
\frac{(M^2)^{i+j}}{(2\pi)^{2n}}
\int\hspace{-2mm}\int\hspace{-1mm}\frac{d^nk_1d^nk_2}
{(k_1^2+i0^+)^{1+i}(k_2^2+i0^+)^{1+j}[(p+k_1+k_2)^2-m^2+i0^+]}.
\end{equation}
   The integrals in Eq.~(\ref{F}) can be written as \cite{Gegelia:zz}
\begin{eqnarray}\label{Fcoeff}
&&\int\hspace{-2mm}\int\frac{d^nk_1d^nk_2}
{(k_1^2+i0^+)^{1+i}(k_2^2+i0^+)^{1+j}[(p+k_1+k_2)^2-m^2+i0^+]}=\nonumber\\
&&\sum_{l=0}^{\infty}(p^2-m^2)^l
f_{ij,l}^{(1)}(m^2,n)+(p^2-m^2)^{2n-5-2i-2j}\sum_{l=0}^{\infty}
(p^2-m^2)^lf_{ij,l}^{(2)}(m^2,n).
\end{eqnarray}
   Inserting Eq.~(\ref{Fcoeff}) into Eq.~(\ref{F}) and taking into
account that $p^2-m^2$ counts as order $Q$, we see that the part
which is proportional to noninteger powers of $p^2-m^2$ (for
noninteger $n$) is of order $Q^{2n-5}$ and therefore satisfies the
power counting.
   The other part is a Taylor expansion in $M^2$ and
$(p^2-m^2)$ and contains terms which violate the power counting.
   This Taylor series can be obtained by formally expanding the integrand
of the original integral of Eq.~(\ref{orint}) in $M^2$ and
$p^2-m^2$ \cite{Fuchs:2003qc} and interchanging summation and
integration.
   Those terms in the Taylor series which do not violate the power counting
contain IR divergences.
   In the IR renormalization, all terms of the so
obtained Taylor series must be subtracted from the original integral.
   [These subtractions are generated by the counterterm diagrams of
Fig.~\ref{fse:fig} (c)].
   However, before the subtraction all IR divergences need to be removed
from the subtraction terms \cite{Schindler:2003xv}.
   This is necessary to ensure that the subtracted expression does not
contain IR divergences.
   In the following, the omission of IR divergences from the subtraction
terms in the IR renormalization is always implied.
   In the EOMS scheme we only need to subtract
those terms which violate the power counting (these terms are free of
infrared divergences)---in the present case all terms of order
$Q^2$ or less \cite{Fuchs:2003qc}.
   Up to and including order $Q^2$ the subtraction terms of the EOMS
renormalization scheme are given by
\begin{equation}
\label{eomsana}
\Delta F^{\rm EOMS}=\Delta_{0,0}+\Delta_{1,0}(p^2-m^2)+\Delta_{2,0}(p^2-m^2)^2
+\Delta_{0,1}M^2,
\end{equation}
where the coefficients are given by
\begin{eqnarray*}
\Delta_{0,0}&=&
-\frac{g^2(m^2)^{n-3}}{2(4\pi)^n}
\frac{\Gamma(3-n)\Gamma(n/2-1)\Gamma(2-n/2)\Gamma(n/2-1)\Gamma(2n-5)}
{\Gamma(3n/2-3)\Gamma(n-2)},\\
\Delta_{1,0}&=&-\frac{g^2(m^2)^{n-4}}{2(4\pi)^n}
\frac{\Gamma(4-n)\Gamma(n/2-1)\Gamma(3-n/2)\Gamma(n/2-1)\Gamma(2n-6)}
{\Gamma(3n/2-6)\Gamma(n-2)},\\
\Delta_{2,0}&=&-\frac{g^2(m^2)^{n-5}}{4(4\pi)^n}
\frac{\Gamma(5-n)\Gamma(n/2-1)\Gamma(4-n/2)\Gamma(n/2-1)\Gamma(2n-7)}
{\Gamma(3n/2-3)\Gamma(n-2)},\\
\Delta_{2,0}&=&\frac{g^2(m^2)^{n-4}}{2(4\pi)^n}
\frac{\Gamma(4-n)\Gamma(n/2-2)\Gamma(3-n/2)\Gamma(n/2-2)\Gamma(2n-7)}
{\Gamma(3n/2-4)\Gamma(n-4)}.
\end{eqnarray*}
   The corresponding subtraction terms of the IR renormalization scheme
read\footnote{In the IR renormalization there is an infinite
number of subtraction terms which contribute in higher orders; in
the following, however, only terms up to the order of the given
calculation will be explicitly shown.}
\begin{equation}
\label{irana}
\Delta F^{\rm IR}=\Delta F^{\rm EOMS}+\delta_{3,0} (p^2-m^2)^3
+\delta_{1,1}(p^2-m^2)M^2
\end{equation}
with
\begin{eqnarray*}
\delta_{3,0}&=&
-\frac{g^2(m^2)^{n-6}}{12(4\pi)^n}
\frac{\Gamma(6-n)\Gamma(n/2-1)\Gamma(5-n/2)\Gamma(n/2-1)\Gamma(2n-8)}
{\Gamma(3n/2-3)\Gamma(n-2)},\\
\delta_{1,1}&=&
\frac{g^2(m^2)^{n-5}}{2(4\pi)^n}
\frac{\Gamma(5-n)\Gamma(n/2-2)\Gamma(4-n/2)\Gamma(n/2-2)\Gamma(2n-8)}
{\Gamma(3n/2-4)\Gamma(n-4)}.
\end{eqnarray*}

\begin{figure}
\epsfig{file=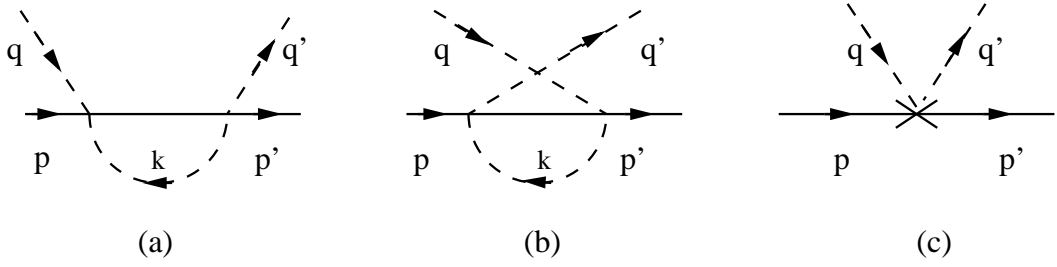,width=14cm} \caption[]{\label{scatt:fig}
Diagrams contributing to $\pi\Psi$ scattering.}
\end{figure}

   Next, let us investigate $G(p^2,m^2,M^2,n)$ which is identified from
Eq.~(\ref{orint}) by rescaling $k_1\mapsto Mk_1, k_2\mapsto k_2$
and $k_1\mapsto k_1, k_2\mapsto Mk_2$.
   Both cases produce an overall factor of $M^{n-2}$.
   One then expands the remaining integrands in $M$, interchanges summation
and integration, and adds up the two contributions.
   Since the integral in Eq.~(\ref{orint}) is invariant under the interchange
of $k_1$ and $k_2$, we only need to perform one of the above manipulations and
then multiply the result with a factor of $2$.
   As a result we obtain
\begin{equation}
\label{G}
G(p^2,m^2,M^2,n)=-\frac{g^2}{(2\pi)^{2n}}\sum_{i,j=0}^{\infty}
\sum_{a=0}^{j}\sum_{b=0}^{a}
(-1)^j2^{j-b}
\left(\begin{array}{c}j\\a\end{array}\right)
\left(\begin{array}{c}a\\b\end{array}\right)
M^{2i+j+b}
I_{ij,ab}(p^2,m^2,n),
\end{equation}
where
\begin{displaymath}
\left(\begin{array}{c}r\\s\end{array}\right)
=\frac{r!}{s!(r-s)!}
\end{displaymath}
is a binomial coefficient and the integrals $I_{ij,ab}(p^2,m^2,n)$ are defined
as
\begin{equation}
\label{loopint}
I_{ij,ab}(p^2,m^2,n)=\int\hspace{-2mm}\int
\frac{d^nk_1d^nk_2\,(p\cdot k_1)^{j-a}(k_1\cdot k_2)^{a-b}(k_1^2)^b}
{(k_1^2-1+i0^+){(k_2^2+i0^+)}^{1+i}{[(p+k_2)^2-m^2+i0^+]}^{1+j}}.
\end{equation}
   For $j+b$ odd, the loop integral of Eq.~(\ref{loopint}) vanishes,
because in that case the integrand is an odd function of $k_1$.
   From this one concludes that the nonvanishing terms in Eq.~(\ref{G}) are
proportional to nonnegative integer powers of $M^2$.
   Using a dimensional counting analysis \cite{Gegelia:zz},
it can be shown that $I_{ij,ab}(p^2,m^2,n)$ is of the form
\begin{equation}\label{loopanalysis}
I_{ij,ab}(p^2,m^2,n)=\sum_{l=0}^{\infty}(p^2-m^2)^lg_{ij,ab,l}^{(1)}(m^2,n)
+(p^2-m^2)^{n+a-b-2i-j-3}\sum_{l=0}^{\infty}(p^2-m^2)^l
g_{ij,ab,l}^{(2)}(m^2,n).
\end{equation}
   Combining this result with Eqs.~(\ref{intsplit}) and (\ref{G}), we
see that the term nonanalytic in $p^2-m^2$ is of order
$Q^{2n-5+a}~(a\ge0)$ and therefore does not violate the power
counting.
   On the other hand, the first part of Eq.~(\ref{loopanalysis}) contains
terms that are analytic in $p^2-m^2$ which, when combined with
Eqs.~(\ref{intsplit}) and (\ref{G}), give rise to contributions which
are nonanalytic in $M$:
\begin{equation}\label{G1}
M^{n-2}G^{(1)}=-g^2M^{n-2}\frac{m^{n-4}\Gamma(2-n/2)}{(4\pi)^{n/2}(n-3)}
\left[1-\frac{1}{2m^2}(p^2-m^2)\right]\frac{\Gamma(1-n/2)}{(4\pi)^{n/2}}
+\cdots.
\end{equation}
   Note that the first term in Eq.~(\ref{G1}) violates the power counting
and, since it is nonanalytic in $M$, cannot be directly absorbed
by a counterterm.
   In solving this apparent puzzle, one has to keep in mind that, in order to
consistently renormalize the diagram of Fig.~\ref{fse:fig} (a),
all relevant contributions from counterterms in $\mathcal{L}_1$ of
Eq.~(\ref{lagrange}) must be taken into account.
   In the present case, these countertems originate in the renormalization of
the one-loop diagrams of Fig.~\ref{scatt:fig} (a) and (b) with the
corresponding expressions
\begin{equation}\label{sub1}
\mathcal{M}_{2a}=
g^2\int\frac{d^nk}{(2\pi)^n}\frac{1}{(k^2-M^2+i0^+)[(k+p+q)^2-m^2+i0^+]}
\end{equation}
and
\begin{equation}\label{sub2}
\mathcal{M}_{2b}=
g^2\int\frac{d^nk}{(2\pi)^n}\frac{1}{(k^2-M^2+i0^+)[(k+p-q')^2-m^2+i0^+]},
\end{equation}
respectively.
   Both diagrams are assigned the order $Q^{n-3}$.
   Evaluating Eqs.\ (\ref{sub1}) and (\ref{sub2}) we find that
$\mathcal{M}_{2a}$ and $\mathcal{M}_{2b}$ contain contributions
that violate the power counting.
   To renormalize these diagrams we follow Refs.\
\cite{Becher:1999he,Fuchs:2003qc} and, up to order $Q$, find the subtraction
terms as
\begin{equation}\label{deleoms}
\Delta\mathcal{M}_{2a}^{\rm EOMS}+\Delta\mathcal{M}_{2b}^{\rm EOMS}
=2ig^2\lambda(m,n)
\end{equation}
 and
\begin{equation}\label{delir}
\Delta\mathcal{M}_{2a}^{\rm IR}+\Delta\mathcal{M}_{2b}^{\rm IR}
=ig^2\lambda(m,n)
\left\{2-\frac{1}{2m^2}[(p+q)^2+(p-q')^2-2m^2]\right\},
\end{equation}
where
\begin{equation}
\label{lambdadef}
\lambda(m,n)=\frac{m^{n-4}\Gamma(2-n/2)}{(4\pi)^{n/2}(n-3)}.
\end{equation}
   The counterterms are then given by
\begin{equation}\label{CTeoms}
\mathcal{L}_{CT}^{\rm EOMS}=
-\frac{1}{2}g^2\lambda(m,n)\pi^2\Psi^2
\end{equation}
for the EOMS scheme and by
\begin{equation}
\label{CTir}
\mathcal{L}_{CT}^{\rm IR}
=-\frac{1}{2}g^2\lambda(m,n)
\left[\pi^2\Psi^2
+\frac{1}{2m^2}\left(\pi^2\Psi\partial_{\mu}\partial^{\mu}\Psi+
2\pi\partial_{\mu}\pi\Psi\partial^{\mu}\Psi
+\pi\partial_{\mu}\partial^{\mu}\pi\Psi^2+m^2\pi^2\Psi^2\right)\right]
\end{equation}
in the IR renormalization.
   In a two-loop calculation these counterterms give a contribution to the
self-energy as shown in Fig.~\ref{fse:fig} (b).
    The corresponding expressions read
\begin{equation}
\label{CTselfeoms}
-i\Sigma_{CT}^{\rm EOMS}=-\frac{1}{2}\,i\int\frac{d^nk}{(2\pi)^n}
\left(\Delta\mathcal{M}_{2a}^{\rm EOMS}
+\Delta\mathcal{M}_{2b}^{\rm EOMS}\right)
\frac{1}{k^2-M^2+i0^+}
=-ig^2\lambda(m,n)\,I_{\pi},
\end{equation}
and
\begin{eqnarray}\label{CTselfir}
-i\Sigma_{CT}^{\rm IR}&=&-\frac{1}{2}\,i\int\frac{d^nk}{(2\pi)^n}
\left(\Delta\mathcal{M}_{2a}^{\rm IR}+\Delta\mathcal{M}_{2b}^{\rm IR}\right)
\frac{1}{k^2-M^2+i0^+}\nonumber\\
&=&-ig^2\lambda(m,n)\left[1-\frac{p^2-m^2}{2m^2}-\frac{M^2}{2m^2}
\right]I_{\pi},
\end{eqnarray}
where
\begin{equation}\label{Ipi}
I_{\pi}=i\int\frac{d^nk}{(2\pi)^n}\frac{1}{k^2-M^2+i0^+}\sim Q^2.
\end{equation}
   In fact, the last term in Eq. (\ref{CTselfir}) is of order $Q^4$,
\begin{equation}\label{Msquared}
M^2I_{\pi}\sim Q^4,
\end{equation}
and can therefore be neglected in a calculation up to and including
order $Q^3$.
   Noting that
\begin{equation}\label{Ipi2}
I_{\pi}=M^{n-2}\frac{\Gamma(1-n/2)}{(4\pi)^{n/2}},
\end{equation}
we see that in the IR renormalization the self-energy contribution of
Eq.~(\ref{CTselfir}), which stems from the renormalization of
one-loop diagrams, exactly cancels the contributions of
Eq.~(\ref{G1}) which are explicitly shown, including the part
which violates the power counting.
   In the EOMS scheme no terms that satisfy the power counting are subtracted.

   Let us finally consider the last term of Eq.~(\ref{intsplit}).
   The function $H(p^2,m^2,M^2,n)$ is obtained from Eq.~(\ref{orint}) by
simultaneously rescaling $k_1\mapsto Mk_1$ and $k_2\mapsto Mk_2$, extracting
a factor of $M^{2n-4}$, expanding the remaining integrand
in $M$, and interchanging integration and summation, yielding
\begin{eqnarray}
\label{H}
H(p^2,m^2,M^2,n)&=&-\frac{g^2}{2(2\pi)^{2n}}
\frac{1}{p^2-m^2}\sum_{i=0}^{\infty}
(-1)^i\left(\frac{M}{p^2-m^2}\right)^i\nonumber\\
&&\times \sum_{j=0}^i
\left(
\begin{array}{c}
i\\
j
\end{array}
\right)
\int\hspace{-2mm}\int d^nk_1d^nk_2
\frac{[2p\cdot(k_1+k_2)]^{i-j}[M(k_1+k_2)^2]^j}{(k_1^2-1+i0^+)
(k_2^2-1+i0^+)}.
\end{eqnarray}
   Noting that $M/(p^2-m^2)\sim Q^0$, it is easy to see that
Eq.~(\ref{H}) in combination with the factor $M^{2n-4}$ of
Eq.~(\ref{intsplit}) satisfies the power counting.
   Furthermore, only nonnegative integer powers of $M^2$ survive in $H$.

   Combining the results above, all terms violating the power counting
are canceled in the sum of the diagrams in Fig.\ \ref{fse:fig} in both
the IR and the EOMS renormalization.
   Finally, we have also studied the case where first $p^2-m^2\rightarrow0$
and then $M\rightarrow0$, and have explicitly verified that the
renormalization procedure remains exactly the same, i.e., the
counterterms are the same and the renormalized diagram satisfies
the power counting.

\section{Conclusion\label{conclusion}}
   In conclusion, we have demonstrated that the application of both the
reformulated IR renormalization and the EOMS scheme to two-loop
diagrams leads to a consistent power counting.
   In this context, the subtraction of one-loop sub-diagrams plays an
important role in the renormalization of two-loop diagrams.
   The procedure can be applied iteratively to multiloop diagrams.
   Calculations using the Lagrangian of baryon chiral perturbation theory are
more involved, but the general features of the renormalization program
do not change.

\acknowledgments
J.~Gegelia acknowledges the support of the Alexander von Humboldt Foundation.

\section{appendix\label{appendix}}

   In this appendix we provide an illustration of the dimensional counting
analysis of Ref.~\cite{Gegelia:zz} in terms of a specific example.
   To that end let us consider the integral of Eq.~(\ref{sub1}) for $q=0$:
\begin{equation}
I_{\pi\Psi}(0,p)=
\frac{i}{(2\pi)^n}\int\frac{d^nk}{(k^2-M^2+i0^+)[(k+p)^2-m^2+i0^+]}.
\end{equation}
   One would like to know how the integral behaves for small values of
$M$ and/or $p^2-m^2$ as a function of $n$.
   If we consider, for fixed $p^2\neq m^2$, the limit $M\to 0$, the
integral $I_{\pi\Psi}(0,p)$ can be represented as
\begin{equation}
I_{\pi\Psi}(0,p)=\sum_i M^{\beta_i} F_i(p^2,m^2,M^2,n),
\label{ippsiexp}
\end{equation}
   where the functions $F_i$ are analytic in $M^2$ and are obtained
as follows.
   First, one rewrites the variable of the loop integration as
$k=M^{a_i}\tilde k$, where $a_i$ is an arbitrary nonnegative real
number.
   Next, one isolates the overall factor of $M^{\beta_i}$ so
that the remaining integrand can be expanded in positive powers of
$M$ and interchanges the integration and summation.
   The resulting series represents the expansion of $F_i(p^2,m^2,M^2,n)$ in
powers of $M^2$.
   The summation over $i$ includes all values of $i$ for which
$F_i$ is nontrivial.
   To be specific, we obtain for $I_{\pi\Psi}(0,p)$:
\begin{equation}
I_{\pi\Psi}(0,p)=\frac{i}{(2\pi)^n}\int\frac{M^{n a_i} d^n\tilde
k}{[\tilde k^2 M^{2 a_i} -M^2+i0^+][\tilde k^2 M^{2 a_i}+2 p\cdot
\tilde k M^{a_i}+ p^2-m^2+i0^+]}. \label{ippsichv}
\end{equation}
   From Eq.~(\ref{ippsichv}) we see that the second propagator does
not contribute to the overall factor $M^{\beta_i}$ for any $a_i$
and has to be expanded in (positive) powers of $(\tilde k^2 M^{2a_i}
+2 p\cdot \tilde k M^{a_i})$ unless $a_i=0$.
   For $0<a_i<1$, we rewrite the first propagator as
\begin{equation}
\frac{1}{M^{2 a_i}} \ \frac{1}{(\tilde k^2-M^{2-2 a_i}+i 0^+) }
\label{1pr}
\end{equation}
and expand the second factor in Eq.~(\ref{1pr}) in positive powers
of $M^{2-2 a_i}$.
   On the other hand, if $1<a_i$ we write the first propagator as
\begin{equation}
\frac{1}{M^{2}} \ \frac{1}{(\tilde k^2 M^{2 a_i-2}-1+i 0^+) }
\label{1pr2}
\end{equation}
and expand the second factor in Eq.~(\ref{1pr2}) in positive
powers of $M^{2 a_i-2}$.
    In both cases one obtains integrals of the type
$\int d^n\tilde k \ \tilde k^{\alpha}$ as the coefficients of the
expansion.
   However, such integrals vanish in dimensional regularization.
Therefore the only nontrivial terms in the sum of Eq.~(\ref{ippsiexp})
correspond to either $a_i=0$ or $a_i=1$.
   Thus we obtain
\begin{equation}
I_{\pi\Psi}(0,p)=I_{\pi\Psi}^{(0)}(0,p)+I_{\pi\Psi}^{(1)}(0,p),
\label{ippsidecomp}
\end{equation}
where
\begin{equation}
I_{\pi\Psi}^{(0)}(0,p)=\frac{i}{(2\pi)^n}\sum_{i=0}^{\infty}
\left( M^2\right)^i \int \frac{d^nk}{[k^2 +i0^+]^{1+i}[k^2 +2
p\cdot k+ p^2-m^2+i0^+]}, \label{ippsi2}
\end{equation}
and
\begin{equation}
I_{\pi\Psi}^{(1)}(0,p)=\frac{i}{(2\pi)^n}\sum_{i=0}^{\infty}
\frac{M^{n-2}(-1)^i}{p^2-m^2} \left(\frac{M}{p^2-m^2}\right)^i
\int\frac{d^n\tilde k \left( \tilde k^2 M+2 p\cdot\tilde
k\right)^i}{[\tilde k^2-1+i0^+]}. \label{ippsi1}
\end{equation}
   On the other hand, the integral $I_{\pi\Psi}(0,p)$ can be
calculated directly without applying the dimensional counting
technique.
   Doing so and comparing with the results given above one
sees that the dimensional counting method leads to the correct
expressions.

   While the loop integrals of Eq.~(\ref{ippsi1}) have a simple
analytic structure in $p^2-m^2$, the technique sketched above can
be directly applied to the loop integrals of Eq.~(\ref{ippsi2}) when
$p^2-m^2\to 0$, now using the change of variable $k=(p^2-m^2)^{c_j}
\ \tilde k$ with arbitrary nonnegative real numbers $c_j$.

\newpage

\end{document}